\documentclass[fleqn,usenatbib]{mnras}

\usepackage{newtxtext,newtxmath}
\usepackage{color}

\usepackage[T1]{fontenc}
\usepackage{ae,aecompl}

\usepackage{graphicx}	
\usepackage{amsmath}	
\usepackage{amssymb}	
\usepackage{color}

\title[On the PPI in TDEs]{On the Papaloizou-Pringle instability in tidal disruption events}

\author[R. Nealon et al.]{
Rebecca Nealon,$^{1,2}$\thanks{E-mail: rebecca.nealon@leicester.ac.uk}
Daniel J. Price,$^{1}$
Cl\'{e}ment Bonnerot$^{3}$
and Giuseppe Lodato$^{4}$
\\
$^{1}$School of Physics and Astronomy, Monash University, Clayton, Victoria 3800, Australia\\
$^{2}$Department of Physics and Astronomy, University of Leicester, Leicester, LE1 7RH, UK\\
$^{3}$Leiden Observatory, Leiden University, PO Box 9513, 2300 RA, Leiden, the Netherlands\\
$^{4}$Dipartimento di Fisica, Universit\`{a} Degli Studi di Milano, Via Celoria, 16, Milano, 20133, Italy
}

\date{Accepted XXX. Received YYY; in original form ZZZ}

\pubyear{2017}

\begin{document}
\label{firstpage}
\pagerange{\pageref{firstpage}--\pageref{lastpage}}
\maketitle

\begin{abstract}
We demonstrate that the compact, thick disc formed in a tidal disruption event may be unstable to non-axisymmetric perturbations in the form of the Papaloizou-Pringle instability. We show this can lead to rapid redistribution of angular momentum that can be parameterised in terms of an effective Shakura-Sunyaev $\alpha$ parameter. For remnants that have initially weak magnetic fields, this may be responsible for driving mass accretion prior to the onset of the magneto-rotational instability. For tidal disruptions around a $10^6$ M$_{\odot}$ black hole, the measured accretion rate is super-Eddington but is not sustainable over many orbits. We thus identify a method by which the torus formed in tidal disruption event may be significantly accreted before the magneto-rotational instability is established.
\end{abstract}

\begin{keywords}
accretion, accretion discs --- black hole physics --- hydrodynamics
\end{keywords}



\section{Introduction}
As a star wanders inside the tidal radius of a black hole, it is disrupted by the black hole's tidal forces and the stellar material is separated in a tidal disruption event (TDE). Roughly half the star remains bound to the black hole, with the remnants of the star maintaining their orbits and evolving into a long gas stream, eventually returning to the black hole. In the case that the debris returning to the black hole is able to circularise to form a disc (or a torus if the gas cannot cool efficiently) and viscously accrete faster than the rate at which the debris returns to pericentre, the accretion rate onto the black hole will be determined by the rate of stellar fallback. With these assumptions, \citet{Rees:1988ly} \citep[but later corrected by][]{Phinney:1989gf} predicted the characteristic $t^{-5/3}$ light curve. \citet{Lodato:2009fj} and \citet{Guillochon:2013kx} later showed that this profile is only accurate at late times and is additionally dependent on the initial internal stellar structure.

The rate at which the gas is able to accrete, cool or circularise depends on the interplay between the viscous accretion, radiative cooling and circularisation timescales \citep[see][for a comparison for different cooling efficiencies]{Bonnerot:2017nr}. In order for a geometrically thin disc to form, gas must be able to cool faster than it circularises \citep{Cannizzo:2009rt,Shen:2014vn}. If instead the circularisation timescale is shorter than the cooling timescale, a geometrically thick torus forms \citep[considered by][]{Loeb:1997uq,Coughlin:2014fr,Piran:2015qy}. In either case mass accretion is assumed to be governed by the magneto-rotational instability (MRI) which grows on the orbital timescale. As the initial magnetic field in the remnant is expected to be weak (typically $\sim$1~G for a solar type star), many orbital timescales are required to grow the field to the point where it can transport angular momentum.

The evolution of such geometrically thick remnants may be complicated by the hydrodynamic Papaloizou-Pringle instability \citep[PPI,][]{Papaloizou:1984qe}. This instability arises in thick, compact tori with well defined inner and outer boundaries (i.e. non-accreting) with a shallow specific angular momentum profile. The torus responds to the instability by developing non-axisymmetric density perturbations that orbit with the gas, resulting in the redistribution of the specific angular momentum, accretion at the inner edge of the torus and damping of the instability \citep{Zurek:1986lr}. Due to the initially weak magnetic field in the TDE remnant, it may be that accretion is initiated by the PPI instead of the MRI.

\begin{figure*}
	\includegraphics[width=1.05\textwidth]{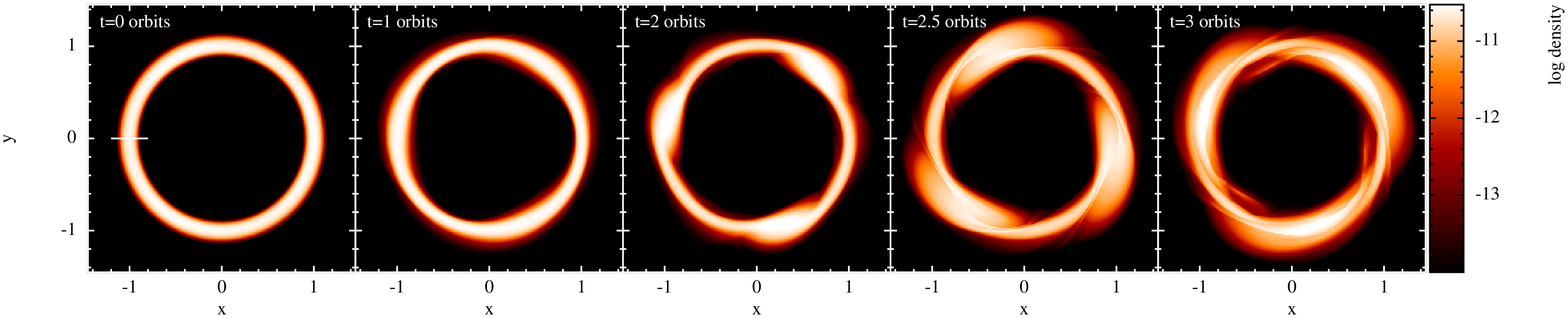}
	\includegraphics[width=1.05\textwidth]{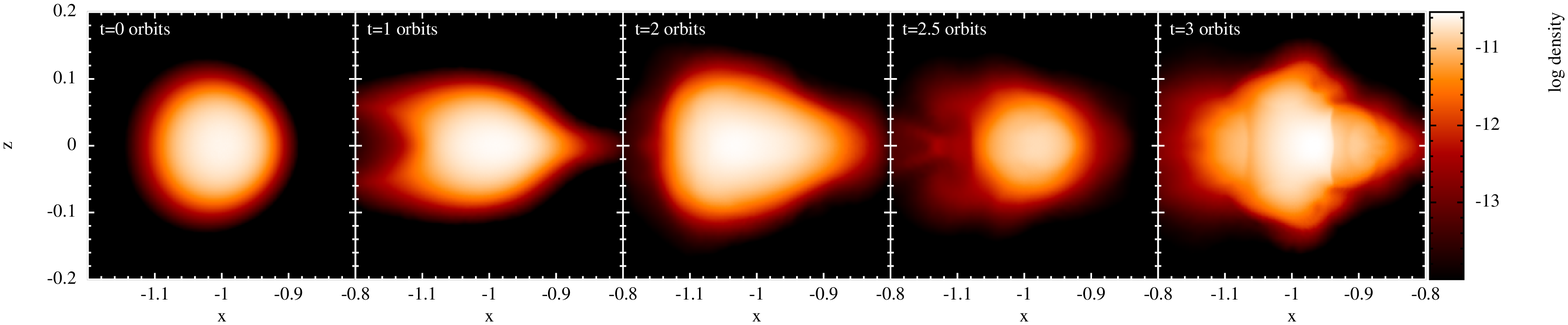}
    \caption{The Papaloizou-Pringle instability in action, showing the development of non-axisymmetric shocks (centre panels) that lead to radial spreading of an initially radially thin torus. Upper panels show the $x$-$y$ cross section and lower the $x$-$z$, with the white line in the top left panel showing the location of the $x$-$z$ cross section. Evolution is from a radially thin torus simulation seeded with an $m=3$ mode. \label{fig:PPI}}
\end{figure*}

Previous work considering these instabilities simultaneously have been inconclusive. \citet{De-Villiers:2002kx} built on work by \citet{Hawley:1991kx}, considering the competition between the MRI and the PPI in a torus around a rotating black hole. In all but one of their simulations, they find that the MRI is able to suppress the PPI by driving accretion at the inner edge. In the latter case they argued that accretion occurred mainly through the midplane such that the inner boundary of the torus was partially well defined, allowing the PPI to develop. \citet{Mewes:2016fk} considered the evolution of PPI-unstable tori around a tilted rotating black hole, finding that self-gravity was critical to the evolution of a torus when the mass of the torus is large relative to the black hole. In their case (not including magnetic fields), the gravitational interaction between the black hole and torus was able to enhance the growth of the dominant PPI mode \citep[see also][]{Krorobkin:2011id}. Most recently, \citet{Bugli:2017kq} conducted a hydrodynamic alongside a full MHD simulation with otherwise identical parameters. They found that the PPI was capable of driving significant accretion of the torus (30\% of the torus mass was accreted in $\sim$18 orbits) but was completely inhibited by the presence of the MRI, even in the case of relatively weak magnetic fields ($\beta \sim$100).

We perform three-dimensional hydrodynamic simulations to investigate the accretion of a tidal disruption remnant using the smoothed particle hydrodynamics (SPH) code \textsc{Phantom} \citep{Phantom}. We begin by simulating a radially narrow torus to quantify the angular momentum transfer that can be driven by the PPI. We then show that a tidal disruption remnant from \citet{Bonnerot:2016qf} is unstable to the PPI. Finally, we simulate a circularised, compact torus representative of the remnant generated by \citet{Bonnerot:2016qf} and quantify the luminosity expected from the PPI driven accretion of this torus. 

\section{The Papaloizou-Pringle instability}
\subsection{Analytical description of the PPI growth}
An analytic description of the Papaloizou-Pringle instability can be found in \cite{Blaes:1986fk} with a simpler analysis presented by \citet{Pringle:2014lr}. They assume a cylindrical flow of incompressible fluid, neglect any self gravity or $z$ dependence and assume an angular velocity profile of the form
\begin{align}
\Omega(R)=\Omega_0 \left(\frac{R_0}{R}\right)^2,
\end{align}
where $R$ is the cylindrical radius and $\Omega_0$ is the angular velocity at a reference radius $R_0$. The frequency $\omega$ of a mode with azimuthal wavenumber $m$ is then a solution of the following equation \citep[e.g.][]{Blaes:1986fk},
\begin{align}
\resizebox{0.5\textwidth}{!}{$\frac{\left(\omega + m \Omega(R_{-})\right)^2 + mg(R_{-})/R_{-}}{\left(\omega + m \Omega(R_{+})\right)^2 + mg(R_{+})/R_{+}} = \left(\frac{R_{+}}{R_{-}}\right)^{2m} \frac{\left(\omega - m \Omega(R_{-})\right)^2 + mg(R_{-})/R_{-}}{\left(\omega - m \Omega(R_{+})\right)^2 + mg(R_{+})/R_{+}},$}
\label{equation:growth_rate}
\end{align}
where the effective gravity is defined according to
\begin{align}
g(R) = \frac{GM R_0}{R^3} \left[1 - \frac{R}{R_0}\right],
\end{align}
with $M$ as the mass of the central object and $R_+$ and $R_-$ denoting the outer and inner edge of the torus, respectively. 

Solutions to Equation~\ref{equation:growth_rate} are determined purely by the choice of the inner and outer radii, demonstrating that the instability is driven by interactions from the boundaries. Two of the four solutions to Equation~\ref{equation:growth_rate} are real, representing stable modes \citep{Blaes:1986fk}. The two remaining solutions correspond to unstable modes, one growing and one decaying. For the growing unstable mode, the growth rate is found from the imaginary component of the frequency, $Im(\omega)$, and depends on the wavenumber $m$.

\subsection{Growth of the PPI in a thin torus}
\label{subsection:PPI_growth}
We start by studying the development of non-axisymmetric perturbations in a radially thin, vertically thick torus. To compare as closely as possible to the description assumed to derive Equation~\ref{equation:growth_rate}, we consider a radially narrow torus with $G=M=1$, $R_0 = 1.0$, $R_+=1.1$ and $R_-=0.9$ (orbital times below are specified at $R_0$). The left-most panels of Figure~\ref{fig:PPI} show the initial torus structure. The density and pressure of the torus are assigned by assuming a polytropic equation of state $P=A\rho^{\gamma}$ and \citep{Papaloizou:1984qe}
\begin{align}
\frac{P}{\rho} = \frac{GM}{(n+1)R_0}\left[\frac{R_0}{r} - \frac{1}{2}\left(\frac{R_0}{r \sin \theta}\right)^2 - \frac{1}{2d}\right],
\end{align}
where the maximum density
\begin{align}
\rho_{\rm max} = \left[\frac{GM}{(n+1)AR_0} \left(\frac{d-1}{2d}\right)\right]^n,
\end{align}
is used to specify $A$. Here $r$ is the radius in spherical coordinates, $R_0$ describes the cylindrical radius of maximum density, $\theta$ is the angle that describes the height out of the plane and $n=(\gamma -1)^{-1}$ is the polytropic index. The factor $d=(R_+ + R_-)/(2R_0)$ determines the profile of the cross section, where values close to unity correspond to a circle. For our radially thin torus we specify an almost circular cross section given by $d=1.01$. In code units, we chose a maximum density $\rho_{\rm max} =  2.5 \times 10^{-9}$ (corresponding to $A=1075$), but the evolution of the torus is independent of this choice. We repeated the simulation using $3.0\times10^5$, $2.3\times10^6$, $1.6\times10^7$ and $1.25\times10^8$ particles. At the lowest resolution the vertical thickness is resolved by approximately two smoothing lengths and in the highest resolution case by ten.

\begin{figure}
	\includegraphics[width=\columnwidth]{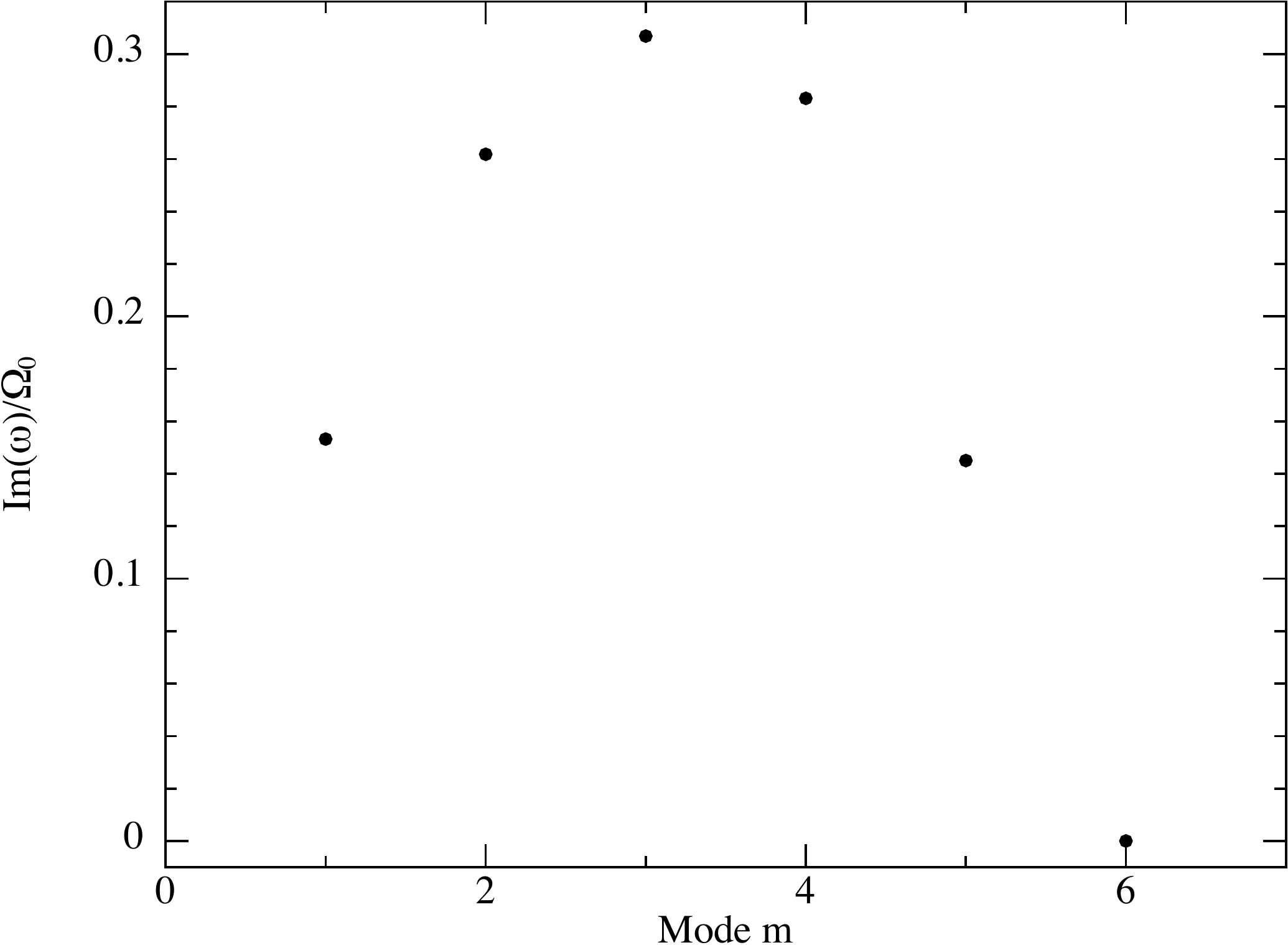}
        \caption{Growth rate calculated from Equation~\ref{equation:growth_rate} for a few azimuthal wavenumbers $m$ when $R_{-}=0.9R_0$, $R_{+}=1.1R_0$. The fastest growing mode ($Im(\omega)\approx 0.3\Omega_0$ for $m=3$) is used to seed the radially narrow torus in Figure~\ref{fig:PPI}.} \label{fig:mode_solution}
\end{figure}

Following \citet{Zurek:1986lr}, the particles in the torus were initially given zero velocity and relaxed in an effective potential that accounted for both the gravitational potential and the pressure forces due to the initial specific angular momentum profile. The torus was allowed to relax in this potential until the potential energy stopped oscillating, corresponding to 5 orbits at $R_0$. Subsequently the particles were given orbital velocities and then seeded with the fastest growing mode. Figure~\ref{fig:mode_solution} shows the growth rates from Equation~\ref{equation:growth_rate} for this choice of $R_-$ and $R_+$, from which we see that $m=3$ is the fastest growing mode. Seeding was achieved with a small azimuthal perturbation in density $\rho$, given by $\rho = \rho_0\left[1 + B \cos (m\phi) \right]$, where $\rho_0$ is the original density, $B$ is the amplitude of the perturbation and $\phi$ is the azimuthal angle. To achieve this the particles were shifted in position by $\delta \phi = -B\sin(m\phi_0)/2$, where $B=0.05$ and $\phi_0$ is the original azimuthal angle \citep[e.g.][]{Price:2007km}. We use a Keplerian gravitational potential with an accretion boundary of $R_{\rm acc}=0.1$ for the lower two resolutions and $R_{\rm acc}=0.2$ for the highest resolution simulation, within which particles are removed from the simulation. The central object has $M=1$ in code units.

The time evolution shown in Figure~\ref{fig:PPI} shows the formation of three over-densities corresponding to the $m=3$ mode, which co-rotate with the fluid at the orbital velocity at $R_0$. After 2.5 orbits (fourth panel from left) these over-densities reach their maximum radial width, with evidence of strong shocks at high resolution. The PPI saturates (fourth panel), and the over-densities become less prominent (fifth panel) as the torus continues to evolve and spread radially. The shocks generated by this instability are particularly clear in the $x$-$z$ cross-section (bottom row) and are confirmed by plotting the divergence of the velocity field.

\begin{figure}
	\includegraphics[width=\columnwidth]{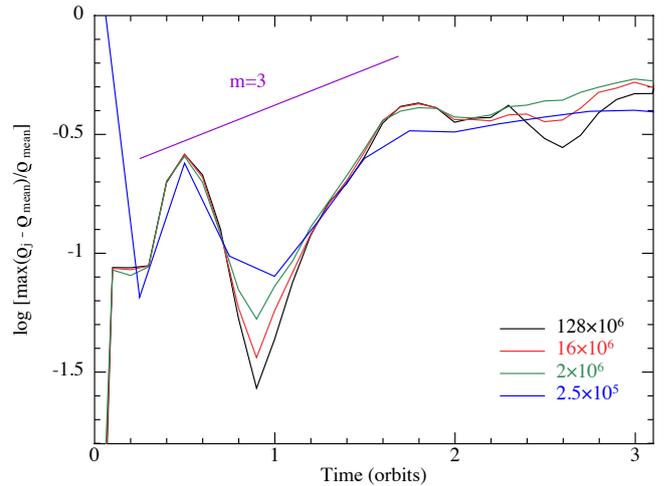}
        \caption{Resolution study showing the amplitude of the non-axisymmetric density perturbation as a function of time from the thin torus simulation. Purple line shows the expected mode growth from Equation~\ref{equation:growth_rate} for the fastest growing ($m=3$) mode. The instability grows on the orbital timescale, saturating after approximately 3 orbits. Comparing the average growth rate in the first two orbits we find numerical convergence in the growth rates.}
    \label{fig:modes}
\end{figure}

To confirm the growth of the density perturbations in our simulation matched Equation~\ref{equation:growth_rate}, we considered particles in the $x$-$y$ cross-section, with $-0.05 < z < 0.05$. We computed the average density, $\rho_{\rm mean}$, from all the particles in this ring. The properties of the particles in the torus were then radially averaged in a method analogous to the azimuthal averaging described in \citet{lodato_2010}, where we divided the torus into $N=45$ azimuthal bins such that each bin represents $\Delta \phi = (360/N)^{\circ}=9^{\circ}$. The deviation from the mean density in each bin $j$ as a function of azimuthal angle was calculated with $(\rho_j - \rho_{\rm mean})/\rho_{\rm mean}$. We measured the density variation at each timestep and identified the maximum as the location of the over-density --- as these co-rotate with the fluid, this occured at a different azimuthal angle at each timestep. Figure~\ref{fig:modes} shows the maximum density variation as a function of time for different numerical resolutions (see legend) compared to the expected growth rate (purple line). We estimated the average growth rate using a least squares fit between 0.1 and 3.0 orbits, measuring $Im(\omega)/\Omega_0=0.5 \pm 0.19$ from the simulations (with the uncertainty derived from the lowest resolution case). Taking into account the assumptions of compressibility and cylindrical flow used to derive Equation~\ref{equation:growth_rate}, we consider our measured growth rate to be consistent with the analytical prediction of $Im(\omega)/\Omega_0\approx0.3$. We additionally visually confirmed that the motion of the over-densities was on the orbital timescale, which was consistent with the analytical prediction of $Re(\omega)/\Omega_0 = m$. The `dip' observed around 1 orbit is caused by mixing of unstable modes with stable oscillations of the torus, and also affected our time averaged growth rate measurement. The amplitude of the perturbations saturated after 2.5 orbits in agreement with Figure~\ref{fig:PPI}. The linear growth phase (first two orbits) is converged for even moderately low resolution, but high resolution is needed for the saturation phase due to the interacting shocks.

\begin{figure}
	\includegraphics[width=\columnwidth]{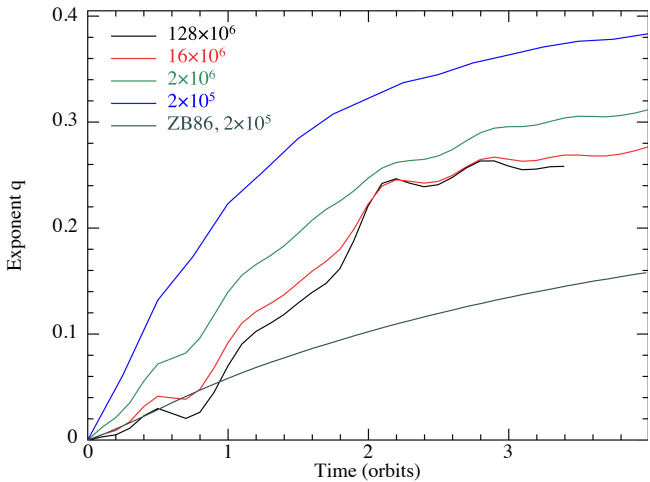} 
    \caption{Evolution of the specific angular momentum profile where $l \propto R^q$ for our four highest resolution simulations and a simulation using the same parameters as \citet{Zurek:1986lr}. In agreement with \citet{Zurek:1986lr}, we find that accretion due to the PPI is equivalent to redistribution of the specific angular momentum profile.}
    \label{fig:q_evolution}
\end{figure}

\section{Does the Papaloizou-Pringle instability lead to angular momentum transport?}
\label{section:viscosity}
Figure~\ref{fig:q_evolution} shows the evolution of the specific angular momentum profile in our radially narrow torus for our four highest resolution simulations, where $l\propto R^q$. To measure $q$ we follow the method outlined in \citet{Zurek:1986lr}, using a least squares fitting procedure to measure the slope of $l$ as a function of $R$. Where they used all the particles inside $R<2$ to measure $q$, we use all the particles in our tori and our uncertainty (included in Figure~\ref{fig:q_evolution}) is derived from the fitting procedure. The profile of the specific angular momentum shown in Figure~\ref{fig:q_evolution} is initially constant (i.e. $q=0$) but development of the PPI over the first three orbits increases $q$. Saturation of the PPI (at roughly 3 orbits, see Figure~\ref{fig:modes}) coincides with the angular momentum settling to a steady profile, with $q \approx 0.25$ for our radially narrow tori.

For comparison, we additionally include a simulation with similar parameters to \citet{Zurek:1986lr} except that we use 100 times more particles (2$\times$10$^5$) to be consistent with our lowest resolution. This simulation was initially unseeded and uses $R_0=0.36$, $A=0.15$, $d=1.36$, finding a lower $q$ value than theirs. As this simulation has the same resolution as our corresponding radially narrow torus (blue line in Figure~\ref{fig:q_evolution}), the different evolution of $q$ should be attributable to the locations of $R_-$ and $R_+$. However, direct comparison with the lower resolution case from \citet{Zurek:1986lr} is not immediately clear. While our simulations identify a strong dependence on resolution for the final $q$ value, this was not observed by \citet{Zurek:1986lr} as they only made use of one resolution. Additionally, \textsc{Phantom} uses a modern viscosity prescription with an $\alpha_{\rm AV}$ viscosity switch and $\beta_{\rm AV}$ viscosity \citep[see][]{Phantom}, neither of which were available to \citet{Zurek:1986lr}. These terms dictate how numerical viscosity is controlled and (particularly at the boundaries) spreading of the torus, ultimately affecting its evolution. Despite these numerical differences, in agreement with \citet{Zurek:1986lr} the initial redistribution of specific angular momentum is rapid and thus likely to be a result of the PPI.

\begin{figure}
	\includegraphics[width=\columnwidth]{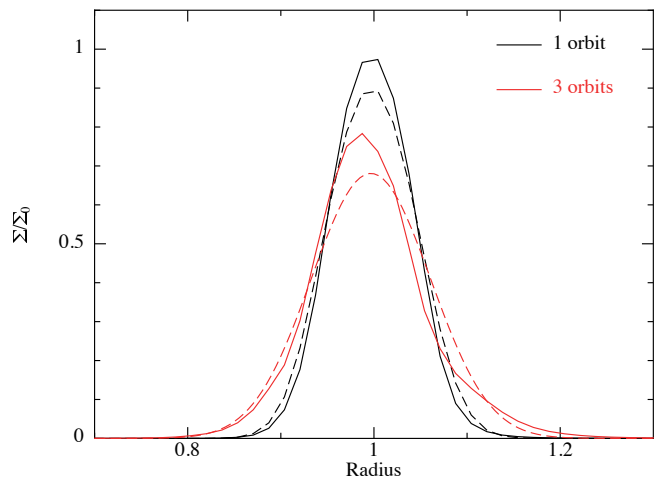}
    \caption{Evolution of the surface density from the thin ring simulation (dashed line), matched to the corresponding evolution of Equation~\ref{equation:propagation} (solid line) at the same time with $q=0.3$. Spreading suggests angular momentum transport may generated by the PPI in a radially thin torus, measured in Figure~\ref{fig:effective_alpha}.}
    \label{fig:ring_spreading}
\end{figure}

\begin{figure}
	\includegraphics[width=\columnwidth]{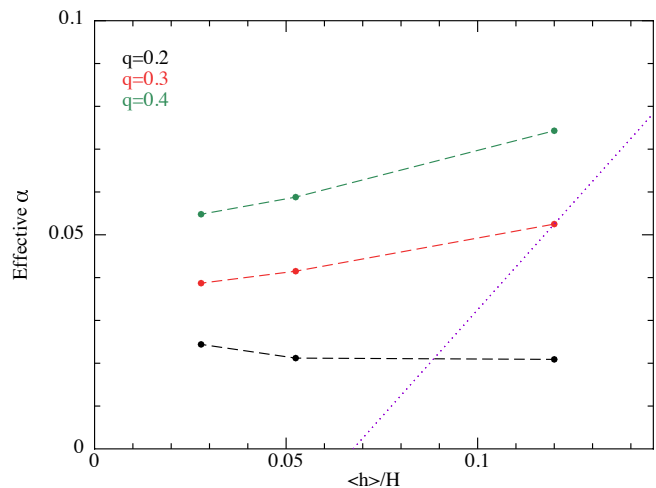} 
    \caption{The `effective $\alpha$' measured from the ring spreading in Figure~\ref{fig:ring_spreading} at our three highest numerical resolutions (increasing from right to left). We find a spreading which is independent of numerical viscosity, with a Shakura-Sunyaev $\alpha \approx 0.04 \pm 0.02$. Here the purple dotted line indicates the expected first-order scaling of the artificial viscosity terms and the spread due to different $q$ values dominates our uncertainty.}
    \label{fig:effective_alpha}
\end{figure}

The radial spreading that occurs in Figure~\ref{fig:PPI} suggests that the PPI may be capable of generating angular momentum transport. Figures~\ref{fig:ring_spreading} and \ref{fig:effective_alpha} quantify this, showing the surface density $\Sigma(r)$ as a function of the spherical radius in the thin ring and the measured `effective $\alpha$' parameter. To measure the viscosity, $\nu$, in the thin torus, which has contributions from both the artificial viscosity in the code and any viscosity generated by the PPI, we use the same technique utilised in \citet{lodato_2010}. That is, we match the surface density evolution in the 3D simulations to the solution of the 1D disc diffusion equation. For the general case where $\Omega = r^{q-2}$, this is given by
\begin{align}
\frac{\partial \Sigma}{\partial t} &= \frac{(2-q)}{q} \frac{1}{r} \frac{\partial}{\partial r} \left[ \frac{1}{\Omega r} \frac{\partial}{\partial r} (\nu \Sigma r^2 \Omega) \right].
\label{equation:propagation}
\end{align}

When $q = 1/2$ this reduces to the expected Keplerian expression \citep[e.g.][]{pringle_1992}. We measure the effective $\alpha$ using a root-finding algorithm, where we minimise the difference between the evolution of $\Sigma$ (modelled by Equation~\ref{equation:propagation}) of the 1D code and our simulation at corresponding times using a finite difference method. We follow the process outlined in Section 4.2.1 of \citet{lodato_2010} in order to quantify the difference between the 1D code and the simulation at a given time step, finding the best fit between surface density in the 1D code and the surface density in the 3D code, using bins within a radial range of $\pm$0.05 around the maximum in the radial surface density profiles.

We set $\Sigma=0$ as the inner boundary condition for the 1D code at $r=0.5$ as in \citet{lodato_2010} and set the same condition far away from the outer edge of the torus at $r=5$. As the $\Sigma$ profile of the relaxed ring cannot be described by a simple power law, the initial $\Sigma$ profile was interpolated directly from the thin torus simulation.  The aspect ratio $H/R$ (where $H$ is measured from the standard deviation of particle position in the $z$ direction) varies between $0.04-0.045$ during the first 4 orbits, so we adopt an average value of $H/R=0.0425$. As the angular momentum profile of our torus changes throughout the simulation (Figure~\ref{fig:q_evolution}), we use $q=0.2$, 0.3 and 0.4. We consider this choice to be the dominant source of uncertainty, as the difference in the measured `effective $\alpha$' for the different $q$ values is greater than the error in our fitting procedure. Figure~\ref{fig:ring_spreading} shows the comparison between the 1D code and the simulation after three orbits of our highest resolution simulation.

Unlike \citet{lodato_2010}, we expect $\alpha$ to vary with time as the PPI develops. Hence we measure $\alpha_n$ at every tenth of an orbit, $t_n$, which represents the average viscosity prior to that time. As the PPI develops the effective viscosity remains roughly constant until 2 orbits, where it steeply increases and reaches a maximum by 2.5 orbits --- when shocks are strongest in the simulation. The effective viscosity then decreases, returning to a constant value by 3 orbits. Figure~\ref{fig:effective_alpha} shows the effective $\alpha$ after the PPI has developed and saturated (i.e. at three orbits) as a function of resolution measured at $R_0$ for different $q$. This effective $\alpha$ has contributions from both the resolution dependent numerical viscosity and the physical viscosity generated by the PPI. Figure~\ref{fig:effective_alpha} demonstrates that the viscosity in the torus is approximately independent of vertical resolution. As the viscosity derived from the higher resolution simulations has a weaker resolution dependence than expected from artificial viscosity only (purple dotted line in Figure~\ref{fig:effective_alpha}), we conclude there is a physical origin for this viscosity that is related to the PPI with an effective $\alpha \approx 0.04 \pm0.02$ for our highest resolution calculation. As shown in \citet{Zurek:1986lr}, the correspondence between this and Figure~\ref{fig:q_evolution} shows that rearranging the specific angular momentum profile is equivalent to transporting angular momentum. These simulations also indicate that we require the ratio of the shell-averaged smoothing length to scale height $\langle h \rangle /H\lesssim 0.1$ in the majority of the torus to guarantee that the rate of spreading in the torus does not change significantly with resolution.

\section{Does the Papaloizou-Pringle instability lead to ballistic accretion?}
We consider whether the PPI allows for ballistic accretion by considering the eccentricity of particles in the torus. Because gas in the torus only travels on Keplerian circular orbits at $R_0$, measuring the eccentricity from the orbital angular momentum is misleading --- it suggests that material in a circular orbit has a non-zero eccentricity. Instead, we consider the motion of the particles that have been accreted throughout the course of the simulation. Tracking these particles we find that material that is accreted starts exclusively at the innermost edge of the torus. This behaviour is in line with what is expected from viscous accretion, confirming that material is not ballistically accreted.

\begin{figure}
	\includegraphics[width=\columnwidth]{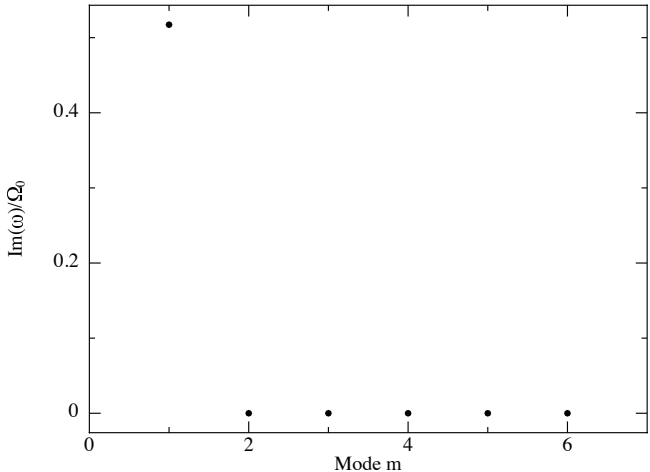}
        \caption{Growth rate calculated from Equation~\ref{equation:growth_rate} as a function of azimuthal wavenumbers $m$ for the tidal remnant torus simulation. In this case $m=1$ is the fastest and only unstable mode, which is seen in Figure~\ref{fig:TDE_evolution}.}
        \label{fig:mode_wide}
\end{figure}

\begin{figure*}
	\includegraphics[width=0.85\textwidth]{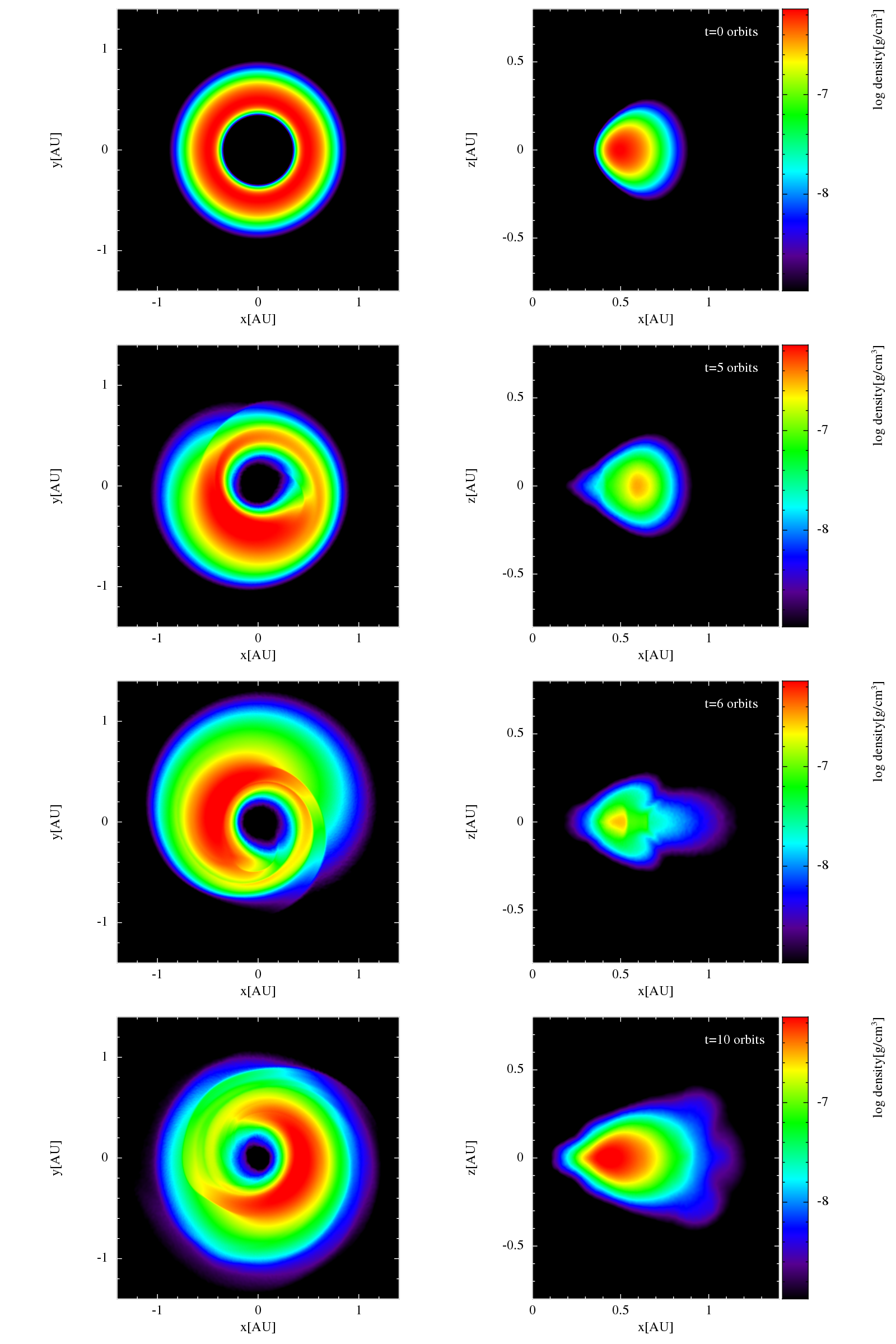}
    \caption{Development of the PPI in a torus that is initially similar to a tidal disruption remnant of a 1~M$_{\odot}$ star around a $10^6$~M$_{\odot}$ black hole. The cross-sectional density evolution from our initial condition until ten orbits (at $R_0 = 0.5$ AU)  is shown on a similar scale to figure 8 of \citet{Bonnerot:2016qf}. As in our previous simulation there is significant spreading in the radial direction due to the PPI.  \label{fig:TDE_evolution}}
\end{figure*}

\section{Are TDE remnants unstable to the PPI?}
Finally, we consider the $1$ M$_{\odot}$ remnant of a TDE around a non-rotating $10^6$ M$_{\odot}$ black hole, calculated by \citet{Bonnerot:2016qf}. This TDE has a penetration factor of $\beta=5$ where $\beta$ is the ratio of the tidal to pericentre radii, an eccentricity of $e=0.8$ and the final evolution after eight orbital periods is shown in their figure 8. We measure the specific angular momentum profile inside $R\sim0.5$ AU (using the method outlined in \S~\ref{section:viscosity}) and find that $0 \lesssim q \lesssim 0.1$. Although the torus formed from the TDE remnant is not yet axisymmetric, the specific angular momentum distribution confirms that if the innermost material circularised it would be susceptible to the PPI.

To consider the subsequent evolution of the torus we do not directly use the torus generated by \citet{Bonnerot:2016qf} because it was found to be already spreading and thus accreting due to artificial viscosity. At higher resolution spreading from artificial viscosity is less pronounced, and so we anticipate that the physical torus will be more radially compact and thus not yet accreting. Taking these considerations into account we consider the evolution of an idealised torus that is radially compact and has similar properties to \citet{Bonnerot:2016qf}, additionally allowing us to investigate the evolution of the PPI in such a torus in a controlled manner. The cross-sectional extent of our torus is also constrained by the torus being in hydrostatic equilibrium. Rather than precisely matching the specific angular momentum measured from the remnant, we adopt constant specific angular momentum (i.e. $q=0$) --- both are unstable to the PPI \citep{Papaloizou:1984qe,Zurek:1986lr}.

We construct a circularised version of the remnant by assuming a torus with a similar cross section and specific angular momentum profile. Similar to our thin torus, we set the particles initially on concentric shells around $R_0=0.5$ AU, using $d=1.15$ and with the number of shells and particles per shell dependent on the resolution (we used $1\times10^5$, $1\times10^6$ and $1\times10^7$ particles, respectively). This torus was relaxed into hydrostatic equilibrium using the relaxing potential as previously, until the potential energy ceased oscillating.

We then set the torus in orbit in a Keplerian potential with a 10$^6$ M$_{\odot}$ central object and added an $m=1$ density perturbation using the method described in Section~\ref{subsection:PPI_growth}. While the $m=1$ mode is the fastest (and only) growing mode for a radially wide torus (see Figure~\ref{fig:mode_wide}), a non-axisymmetric density structure with this mode is already present in the remnant produced by \citet{Bonnerot:2016qf} (their figure 5). Additionally, if simulated with no added perturbation we find that the $m=1$ mode develops. Again we use a Keplerian potential, but here with an accretion boundary at $R_{\rm acc}=0.1$  AU.

Figure~\ref{fig:TDE_evolution} shows the initial density cross section of this torus and after 5, 6 and 10 orbits. Within the first three orbits (measured at $R_0$) an asymmetry develops but is not yet accompanied by appreciable spreading. The instability continues to grow, and by five orbits a shock has developed that extends from the inner to outer edges. By this stage the outer radius has spread by $\approx$10\% at the location of the over-density. The PPI continues to grow and the torus reaches its maximum size by six orbits, when the PPI visually saturates. By seven orbits both the strength of the shock and the asymmetry of the outer edge decrease, although the over-density remains (the partial remnants may be seen in the lower panels of Figure~\ref{fig:TDE_evolution}). The $x$-$z$ plane cross section (right hand side) also shows the effects of the shocks at the outer edge of the torus. While the torus has not spread much in the $z$ direction, it has doubled its radial extent (although the inner edge is constrained by our choice of accretion radius). For a TDE remnant that does circularise, this simulation suggests that the PPI develops within a few orbits, creating over-densities that remain even after the instability has saturated.

\begin{figure}
	\includegraphics[width=\columnwidth]{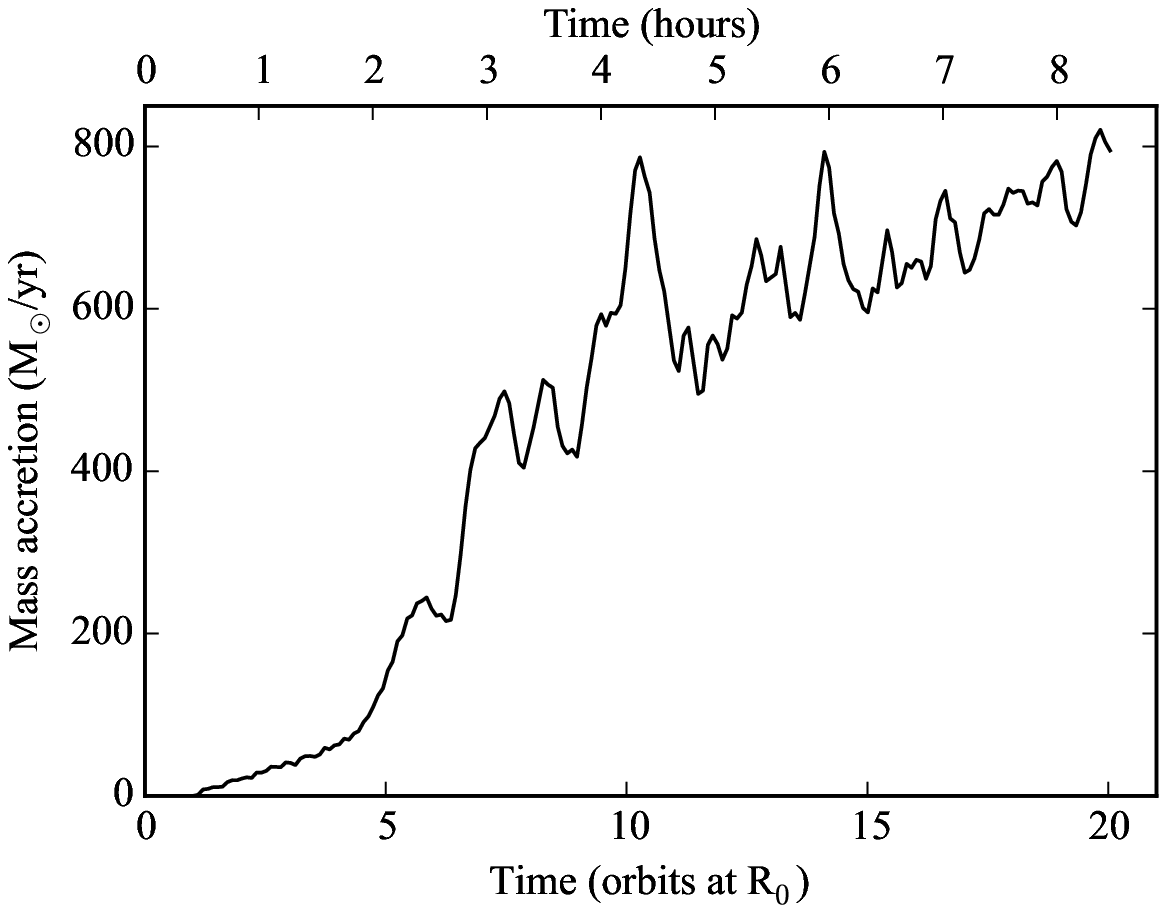}
        \caption{Mass accretion rate as a function of time in the tidal disruption remnant simulation shown in Figure~\ref{fig:TDE_evolution}. Accretion is primarily driven by the growth and saturation of the Papaloizou-Pringle instability in the remnant where the $m=1$ mode dominates. The mass accretion rate is super Eddington ($\dot{M}_{\rm Edd} = (\epsilon/0.1)\times 2.1\times10^{-2}$ M$_{\odot}$/yr).\label{fig:mass_accretion}}
\end{figure}

Figure~\ref{fig:mass_accretion} shows the mass that falls within the accretion boundary of the high resolution TDE simulation over $\sim$ 20 orbits scaled with the black hole and star parameters specified in \citet{Bonnerot:2016qf}. Within the first five orbits this leads to a mass accretion rate of $\sim$100~M$_{\odot}$/yr --- this this is well before the MRI is expected to be able to drive significant mass accretion (however, see Section~\ref{section:five}).

With the average mass accretion rate while the PPI is active but not yet saturated (e.g. around 5 orbits), the luminosity of the accreting material can be estimated using
\begin{equation}
L = \epsilon \dot{M} c^2.
\end{equation}
When the PPI is constraining the accretion rate the luminosity is found to be $L \approx 1.1 \times 10^{48}$ erg/s, assuming $\epsilon=0.1$. This is $\sim$10$^4$ times larger than the Eddington luminosity for such a black hole ($L_{\rm Edd} = 1.28 \times 10^{44}$ erg/s).

\section{Which instability drives initial accretion?}
\label{section:five}
The PPI growth rate for the only unstable mode in the tidal disruption remnant torus is $\sim$0.5$\Omega_0$. As this is slower than the growth rate of the MRI ($0.75\Omega_0$), it would be expected that the faster growing instability would drive the initial accretion of the torus. However, the initial magnetic field in the torus formed from a tidal disruption event is expected to be around the same magnitude as a star, $\sim$1 G. This weak initial field means that the MRI will have to grow for many orbits before accretion can be established. During this time the PPI may be able to saturate and hence drive accretion.

\begin{figure}
	\includegraphics[width=\columnwidth]{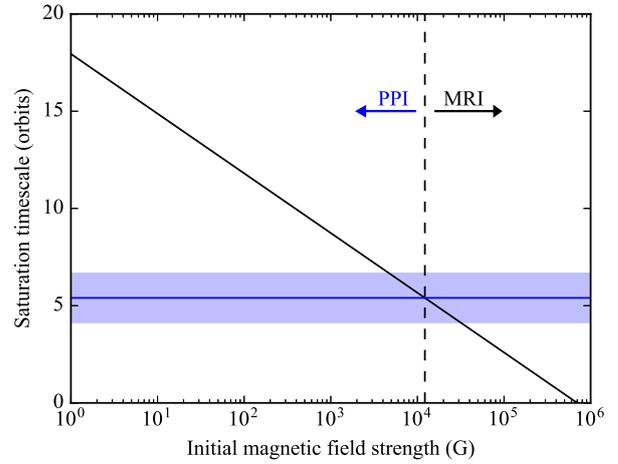}
        \caption{Timescale for saturation as a function of initial magnetic field strength of the Papaloizou-Pringle (blue) and magnetorotational instabilities (black) expected in the tidal disruption remnant shown in Figure~\ref{fig:TDE_evolution}. The arrows indicate the region where accretion is initially dominated by either instability. The PPI should drive accretion before the MRI in tori with reasonable initial magnetic fields (i.e. $\sim$1 G), despite the faster growth rate of the MRI.}
        \label{fig:time_scales}
\end{figure}

We thus consider the saturation timescales of both the PPI and the MRI to be indicative of the timescale for significant angular momentum transport. Figure~\ref{fig:time_scales} shows these timescales for a range of initial magnetic field strengths. We measure the saturation of the PPI in the tidal disruption remnant as the time taken for strong shocks to develop. We consider the equivalent of Figure~\ref{fig:modes} for this torus in our high resolution simulation (with 1$\times 10^7$ particles) and identify when the density variation stops increasing (the equivalent of $\approx$1.8 orbits in Figure~\ref{fig:modes}). The uncertainty in this measurement (represented by the blue shaded region in Figure~\ref{fig:time_scales}) is estimated by repeating this process for the simulation with 1$\times 10^6$ particles, as our lowest resolution simulation does not display any clear saturation. The maximum field strength required to establish the MRI will occur when the Alfv\'{e}n speed and sound speed are comparable. However the MRI can be established at lower field strengths than this, shown recently by \citet{Bugli:2017kq} and in cases where there is zero net flux \citep[e.g.][]{Hawley:2001lr}. As a result, we consider weaker magnetic fields where the ratio between the gas and magnetic pressure $\beta\sim$100. The magnetic field required for saturation in this case is
\begin{align}
B_{\rm sat} \approx \frac{1}{10} H \Omega \sqrt{4\pi \rho},
\label{equation:MRI}
\end{align}
\begin{align}
\resizebox{0.5\textwidth}{!}{$\approx 7.1\times 10^5 {\text{ G}} \left(\frac{H}{4.5\times10^{12} {\text{ cm}}} \right) \left( \frac{\Omega}{5.6\times10^{-4} {\text{ s}}}\right) \left( \frac{\rho}{6.31\times 10^{-7} {\text{ g cm}}^3} \right)^{1/2}. $}\nonumber
\end{align}
Using the maximum density and scale-height from the simulation in Figure~\ref{fig:TDE_evolution} the magnetic field strength required for our torus is $B_{\rm sat}\approx 7.1 \times 10^5$~G. The saturation timescale for the MRI in terms of the initial magnetic field $B_{\rm initial}$ is
\begin{equation}
t_{\rm MRI} = \frac{1}{0.75 \Omega_0} ln\left( \frac{B_{\rm sat}}{B_{\rm initial}} \right).
\label{equation:MRI_time}
\end{equation}

The comparison in Figure~\ref{fig:time_scales} demonstrates that the PPI may drive significant accretion before the MRI for tori susceptible to the PPI with magnetic field strengths below $\sim$10$^4$ G. In the case of the tidal disruption remnant shown in Figure~\ref{fig:TDE_evolution} with an initial magnetic field of 1 G, this corresponds to accretion due to the PPI for $\sim$7 hours before MRI accretion begins ($\sim$18 orbits).

Although weak magnetic fields are considered here, the field strength calculated is still likely to be an overestimate. Accretion that is driven while the MRI is developing but not established may be capable of damping the PPI, as even a low accretion rate is able to prevent this instability. This would cause the PPI to be damped before the field is saturated, decreasing our estimate in Equation~\ref{equation:MRI}. Additionally, Figure~\ref{fig:time_scales} does not take into account the initial eight periods that have already occurred during the original torus formation by \citet{Bonnerot:2016qf}. Although the MRI may begin to develop during this time, because the structure of the tidal disruption remnant is still evolving it is not obvious that the MRI growth rate will be the same as Equation~\ref{equation:MRI_time}. Recent work by \citet{Bugli:2017kq} appears to confirm our findings as an overestimate, however their simulations were initialised with a random perturbation while ours are seeded with the fastest growing mode. In our case this is appropriate (as the initial torus by \citealt{Bonnerot:2016qf} has an existing non-axisymmetric perturbation), but this difference makes direct comparison with \citet{Bugli:2017kq} more difficult. Their work was also limited by not including any explicit magnetic diffusivity, which may be capable of quenching the MRI under certain circumstances \citep[see discussion,][]{Bugli:2017kq}. Despite our overestimation of the field required, the magnetic field strength expected in a star undergoing tidal disruption suggests that the PPI may drive the initial accretion in these tori.

\section{Discussion}
Informed by the resolution study of the thin ring simulation in Figure~\ref{fig:effective_alpha}, we would consider the simulation of the tidal remnant with $\langle h \rangle /H \approx 0.05$ to be resolved. A resolution study (not shown) suggests that the converged mass accretion rate would be $\sim$3 times lower than that shown in Figure~\ref{fig:mass_accretion}, despite Figure~\ref{fig:modes}~and~\ref{fig:effective_alpha} suggesting that the development of the PPI is fully resolved. For the measured super-Eddington luminosity, we expect that strong outflows would develop, either in the form of an expanding Eddington-limited spherical bubble \citep{Loeb:1997uq} or in the form of strong winds \citep{Lodato:2011ij}.

The original simulation by  \citet{Bonnerot:2016qf} assumed an elliptical rather than parabolic orbit such that our mass accretion is likely to be an overestimate. This choice means that the entire star forms into the torus rather than the $\lesssim 50$\% expected from theoretical predictions, making our mass accretion rate roughly an order of magnitude larger than it should be. The accretion timescale is also shorted by the elliptical orbit (by a factor of $\approx 0.03$ for $e=0.8$), further compounding our overestimate. However, even while taking these differences into account (and the lower mass accretion rate discussed above) we still predict a super-Eddington mass accretion rate --- suggesting a high mass accretion rate remains likely for parabolic encounters.

The mass accretion rate from the tidal disruption remnant simulation suggests that accretion associated with the PPI should occur before any MRI driven accretion, with a delay of about $\sim$7 hours in our tidal disruption remnant simulation. Additionally, the density structure that is generated in the torus leads to modulation of the mass accretion rate. In the case that the mass accretion rate corresponds directly to the luminosity or any outflows generated, this would suggest that these features may also initially be modulated.

An interesting corollary of our work is the behaviour of the mass accretion rate after the PPI has saturated. Here we have assumed that the MRI would take over at this point, driving subsequent accretion. In the early evolution of the torus when accretion is driven by the PPI, does the rate of accretion lead to self regulation or self-damping of the instability? As accretion due to numerical viscosity becomes important in our simulations around this point, this is difficult to consider with our current simulations, but may be worth exploring in future work.

\section{Conclusions}
We have investigated the evolution of circularised tori with well defined inner and outer boundaries, similar to the compact discs formed by the tidal disruption of stars around a supermassive black hole. We demonstrate unmagnetised remnants of tidal disruption events are unstable to the $m=1$ mode of the Papaloizou-Pringle instability. We find that this instability drives ring spreading and angular momentum transport that may be parameterised in terms of a Shakura-Sunyaev $\alpha$ viscosity. Finally, the evolution of this instability in a circularised tidal disruption remnant can drive a super-Eddington mass accretion rate prior to the onset of the magneto-rotational instability. As the accretion will be driven by the magneto-rotational instability after about eighteen orbits, these accretion rates are not expected to be sustained for more than a few orbits.

\section*{Acknowledgements}
We thank Yuri Levin and Chris Matzner for useful discussions and the anonymous referee for valuable comments that improved the manuscript. This project has received funding from the European Research Council (ERC) under the European Union's Horizon 2020 research and innovation programme (grant agreement No 681601). DP acknowledges funding from the Australian Research Council via FT130100034 and DP130103078. This work was performed on the gSTAR national facility at Swinburne University of Technology. gSTAR is funded by Swinburne and the Australian Government's Education Investment Fund. We used \textsc{SPLASH} \citep{Price:2007kx} for Figures~\ref{fig:PPI}~to~\ref{fig:TDE_evolution}.




\bibliographystyle{mnras}
\bibliography{master} 


\bsp	
\label{lastpage}
\end{document}